\documentclass[9pt,technote]{IEEEtran} 
\usepackage{amsmath, amssymb, amsthm,mathtools, stmaryrd, bm, gensymb,cite,algpseudocode,float,subfloat}
\usepackage{subfigure,xcolor}
\usepackage{caption}
\usepackage{siunitx}
\usepackage[english,ruled,lined,linesnumbered]{algorithm2e} 
\usepackage[none]{hyphenat}

\def\BibTeX{{\rm B\kern-.05em{\sc i\kern-.025em b}\kern-.08em
    T\kern-.1667em\lower.7ex\hbox{E}\kern-.125emX}}
\usepackage{balance}

\newcommand{\B}[1]{\mathbf{#1}}

\newcommand{\R}[1]{\mathrm{#1}}
\newcommand{\bof}[1]{{\mbox{\boldmath$#1$}}}
\newcommand{\gf}[1]{\textcolor{black}{{#1}}}
\newcommand{\fa}[1]{\textcolor{black}{{#1}}}
\newcommand{\fea}[1]{\textcolor{black}{{#1}}}
\begin{document}
\title{Structured Channel Estimation for RIS-Assisted THz Communications}

\author{Fazal-E-Asim,~\IEEEmembership{Senior~Member,~IEEE}, Bruno Sokal,~\IEEEmembership{Member,~IEEE}, Andr\'e L. F. de Almeida,~\IEEEmembership{Senior~Member,~IEEE}, Behrooz Makki,~\IEEEmembership{Senior~Member,~IEEE}, and G\'abor Fodor,~\IEEEmembership{Senior~Member,~IEEE}
\thanks{This work was partially supported by the Ericsson Research, Sweden, and Ericsson Innovation Center, Brazil, under UFC.51. The authors acknowledge the partial support of Fundação Cearense de Apoio ao Desenvolvimento Científico e Tecnológico (FUNCAP) under grants FC3-00198-00056.01.00/22 and  ITR-0214-00041.01.00/23, the National Institute of Science and Technology (INCT-Signals) sponsored by Brazil's National Council for Scientific and Technological Development (CNPq) under grant 406517/2022-3. This work is also partially supported by CNPq under grants 312491/2020-4 and 443272/2023-9. Also, G. Fodor was supported by the EU Horizon Europe Program, Grant No: 101139176 6G-MUSICAL.

Fazal-E-Asim, Bruno Sokal, and Andr\'e L. F. de Almeida all are with the Wireless Telecommunications Research Group (GTEL), Department
	of Teleinformatics Engineering, Federal University of Ceara, Brazil.
	Behrooz Makki is with Ericsson Research, Ericsson, 417 56 Gothenburg, Sweden.
	G\'abor Fodor is with Ericsson Research, 16480 Stockholm, and also with the Division of Decision and Control, KTH Royal Institute of Technology, 11428 Stockholm, Sweden. 
	e-mail: \{fazalasim, brunosokal, andre\}@gtel.ufc.br. \{behrooz.makki, gabor.fodor\}@ericsson.com.}  }
\maketitle
{\color{black} 
\begin{abstract}
This paper proposes tensor-based channel estimation for reconfigurable intelligent surface (RIS)-assisted communication networks. We exploit the inherent geometrical structure of the Terahertz propagation channel, including the antenna array geometries at the base station, the RIS, and the user equipment to design a tensor-based channel estimator, referred to as the higher-dimensional rank-one approximations (HDR) method. By exploiting the geometrical structure of the combined \gf{base station}-RIS-\gf{user equipment} channel, the proposed HDR estimator recasts parametric channel estimation as a single sixth-order rank-one tensor approximation problem, which can be efficiently solved using \gf{higher-order singular value decomposition} to deliver parallel estimates of \gf{each channel component vector}. Numerical results show that the proposed method provides significantly more accurate parameter estimates than competing state-of-the-art \fa{tensor-based RIS channel estimation}, Khatri-Rao factorization, and \gf{least squares} methods. \fa{For higher-rank channels, the HDR method shows similar spectral efficiency compared to its competitors while having similar computational complexity to the classical least squares estimator.}
\end{abstract}
}
\begin{IEEEkeywords}
Rank-one approximation, Terahertz
(THz) communications, channel parameter estimation, and tensor modeling.
\end{IEEEkeywords}

%
\IEEEpeerreviewmaketitle

\vspace*{-0.3cm}
\section{Introduction}
%
%
%
%
\IEEEPARstart{R}{econfigurable} \gf{intelligent surface (RIS) is a promising candidate for the enhancement of future communication networks.} Moreover, the introduction of millimeter-wave (mmWave) and Terahertz (THz)/(micrometer) bands facilitates the deployment of a massive number of antennas at the base station (BS) and the user equipment (UE). 
To deal with these problems, RISs that help to concentrate energy towards the desired location and minimize the interference elsewhere \gf{is} a promising solution \gf{to boost the signal-to-interference-plus-noise ratio in an energy-efficient fashion}. On the other hand, estimating wireless channels is challenging when dealing with large RIS panels in combination with massive multiple input multiple output (MIMO) infrastructure nodes.

\gf{Acquiring channel state information (CSI)} is especially challenging when passive RISs are deployed, where the burden of pilot transmission, reception, and associated signal processing must be managed by the end nodes of the communication network \cite{Renzo_2020}. 
\gf{Furthermore, with the use of higher frequencies, the communication channel is dominated by a few usable propagation paths, which are more susceptible to blockages that decrease the coverage, throughput, and quality of services provided by the network.} Here, \gf{RISs} can effectively avoid the blockage effects by establishing a virtual path. 

\fea{To summarize, a passive RIS cannot process pilot signals, which indirectly means that channel estimation/parameter estimation must be performed at the end nodes of the wireless network, potentially overloading the chosen node. Secondly, a large panel size usually implies many channel coefficients to be estimated. This may require many pilot resources, affecting the overall system spectral efficiency (SE). Finally, the accuracy of the CSI acquisition is crucial because the gains the RIS provides depend on a sufficiently accurate knowledge of the channel.}

\fea{The paper \cite{Swindlehurst_ALee_2022} briefly explains various channel estimation methods in RIS-assisted communications, considering both unstructured and structured channel models. Algorithms introduced for estimating unstructured channels are simple and easy to implement but require extensively huge training overhead in RIS-based systems, compromising their achievable rates due to the estimation of many channel coefficients. On the other hand, geometric channel models lead to an estimation of fewer channel parameters, and therefore, much smaller training is required for improved estimation performance. These benefits, however, come at the cost of increased algorithmic complexity, model order estimation, and inevitable modeling errors. }

The authors in \cite{Masood_2022} use the inductive matrix completion followed by a root multiple signal classification (MUSIC) algorithm to estimate the respective angle of departure and arrival in RIS networks. \fea{However, the method is only shown for a uniform linear array (ULA), including RIS, with no clear method for angle pairing.} Also, \cite{Wang_Peilan_2020} exploits the sparsity of the mmWave propagation channel to jointly design beamforming and \gf{estimating} the \gf{cascaded} BS-RIS-UE channel. \fea{However, the proposed method is designed for a single-antenna user with a ULA BS. The work in \cite{Muye_2022} designs a joint channel estimation and data detection algorithm for hybrid reconfigurable intelligent surface (HRIS) using orthogonal time frequency space modulation. 
A new transmission structure is formulated so that the partial HRIS elements are alternatively activated. The channel and unknown data symbols are estimated by iteratively executing the message passing and expectation-maximization algorithms.
Therein, a ULA BS and a single-antenna user are assumed, which impedes full exploitation of the channel structure.  } 
The works in \cite{Xiaoling_2023,Zhouyuan_2022} used distributed semi-passive IRS for sensing and passive IRS for communications in an integrated sensing and communication (ISAC) system for single and multi-user cases. The total least squares (LS) estimation of signal parameters via the rotational invariance
technique is used to estimate the angles of arrival effectively. Then, as a second step, the MUSIC algorithm is used to obtain effective pairing. However, single antenna users are considered in these works.

Tensor decompositions have been successfully applied to formulate channel estimation methods for mmWave/massive MIMO systems in different scenarios. In particular, the widespread use of the parallel factor (PARAFAC) tensor decomposition in wireless communications comes from its capability to exploit the intrinsic multilinear structure of signals/channels and its powerful uniqueness properties \cite{Pierre2009tensor}. The authors in \cite{Daniel_2019} combine tensor decomposition and compressive sensing to formulate a sparse channel estimation method for massive MIMO-OFDM (orthogonal frequency division multiplexing) systems. The work \cite{Zilli_2021} proposes a joint hybrid precoder and combiner design for maximizing the achievable sum rate of mmWave MIMO-OFDM systems, where the analog precoder and combiner are designed as a Tucker2 tensor decomposition problem.

       Tensors have recently been exploited to derive channel estimation methods for RIS-assisted communications \cite{Gilderlan_2021,Araujo2023,KhaledArdah_2021,GherekhlooK_2021,Gomes_2023,benicio2023tensor_wcl}. 
      The authors in \cite{Gilderlan_2021} capitalize on the PARAFAC decomposition to formulate an iterative algorithm based on alternating least squares (ALS) \cite{CLdA09} to solve the individual channel estimation problem for RIS-assisted MIMO communications. The work  \cite{Xinran_2022} jointly exploits the sparsity and tensor decomposition structures to improve channel estimation performance for RIS in a multiuser scenario. The work \cite{KhaledArdah_2021} introduces sparsity-structured-based PARAFAC tensor factorization. A two-stage sparse recovery problem is proposed for mmWave channels, referred to as two-stage RIS-aided channel estimation (TRICE). Therein, the first stage provides the directions of departure and arrival estimates, while in the second stage, the cascaded channel is estimated from the previously extracted angular parameters. The work \cite{GherekhlooK_2021} exploits low-rank channels and designs a PARAFAC decomposition, i.e., tensor-based RIS channel estimation (TenRICE), where the channel matrices are estimated using an ALS method. \fea{However, the latter TenRICE method outperforms the TRICE method.}
      
          In \cite{Gomes_2023},  tensor-based channel estimation schemes are proposed for IRS-MIMO systems considering hardware imperfections affecting the phase shifts. The work \cite{benicio2023tensor_wcl} formulates a tensor-based approach to channel estimation and data tracking in RIS-assisted MIMO systems. The reader is referred to \cite{Almeida2016} and \cite{Chen_2021} for overviews on tensor decompositions applied to wireless communications.
      Reference \cite{deAlmeida2007parafac} proposes PARAFAC-based unified tensor modeling for wireless communication and analyses various applications subject to frequency-selective multipath fading, such as blind multiuser equalization. 
      Paper \cite{favier2014overview} provides an overview of constrained PARAFAC models having linear dependencies among columns of the factor matrices of the tensor decomposition.
The work in [11] jointly exploits the sparsity and tensor decomposition structures to improve channel estimation performance for RIS-assisted communication in a multiuser scenario having a single antenna each. \gf{That} work introduces sparsity-structured-based
PARAFAC tensor factorization.

{\color{black} Finally, \cite{Jia_2020} proposes an adaptive grid-matching pursuit estimation algorithm by transforming the estimation problem into a sparse recovery one, while in \cite{Kim_2022} a variational inference-sparse
Bayesian learning estimator is proposed for RIS-assisted networks.} \fea{Most of the previous works either use single antenna users or deploy ULA both at the UE and the BS. Please note that although the proposed work in this paper, assumes a uniform rectangular array (URA) at both the BS and the UE, the proposed method still applies to ULA deployed at both ends.}

In this paper, we exploit the geometrical structure of the channel estimation problem and recast it as a rank-one higher-order tensor approximation problem to estimate the channel parameters \fa{associated with the dominant communication links.} More specifically, we reformulate the BS-RIS-UE channel as \fa{an approximated} sixth-order rank-one channel tensor, which allows us to recast parametric channel estimation as a single rank-one tensor approximation problem, which can be efficiently solved using the higher-order singular value decomposition (HOSVD) algorithm \cite{DeLathauwer2000}. Our proposed higher-dimensional rank-one approximations (HDR) method offers remarkable gains over its competitors in terms of normalized mean square error (NMSE) in the low signal-to-noise-ratio $(\R{SNR})$ regime. Moreover, our proposed HDR method outperforms the 
\gf{Cram\'{e}r-Rao lower bound
(CRLB)} derived in \cite{Gilderlan_2021}, due to its efficient noise rejection by exploiting the rank-one tensor channel structure.
\fea{ The contributions of the present paper are summarized as follows:}
\begin{itemize}
 \item \fea{Exploiting the geometrical structure of the propagation channels, we resort to a tensor modeling formalism to formulate the received pilot signal as a higher-order tensor.}
\item \fea{Using a data reshuffling scheme, we recast the structured channel estimation problem as an HDR problem to estimate the channel
parameters associated with the dominant communication links.}
\item \fea{Analyzing the performance of the proposed HDR method for several scenarios and comparing it with the classical LS method \cite{Gilderlan_2021} and the theoretical CRLB \cite{Gilderlan_2021} for unstructured channels. }
\item \fea{Comparing our solution with state-of-the-art competing methods such as Khatri-Rao factorization (KRF) \cite{Gilderlan_2021} and TenRICE \cite{GherekhlooK_2021}. }
     \end{itemize}
\textbf{\textit{{Notation:}}} Scalars are denoted by lower-case italic letters $a$, vectors by bold lower-case italic letters $\bof{a}$, matrices by bold upper-case italic letters $\bof{A}$, tensors are defined by calligraphic upper-case letters $\mathcal{A}$. $\bof{A}^\R{T}$,$\bof{A}^\R{\ast}$,$\bof{A}^\R{H}$ stand for transpose, conjugate and Hermitian of $\bof{A}$. The operators $\otimes$, $\diamond$, $\circ$, and $\odot$ define the Kronecker, the Khatri-Rao, the outer product, and the Hadamard (element-wise) product, respectively. $\R{diag}\{\B{a}\}$ represents the square diagonal matrix with elements of vector $\B{a}$ across the main diagonal. $\R{vec}\left\{.\right\}$ vectorizes an $I \times J$ matrix argument, while $\R{unvec}_{(I \times J)}\left\{.\right\}$ does the opposite operation. $\mathbb{E}[\cdot]$ is expectation operator.
For an $N$th order tensor $\mathcal{Y} \in \mathbb{C}^{L_1\times\dots\times L_N}$, the $n$-mode unfolding of $\mathcal{Y}$ is the matrix $\left[
	\mathcal{Y}\right]_{(n)} = \mathbb{C}^{L_n \times L_1\dots L_{n-1}L_{n+1}\dots L_N}$.
\vspace*{-0.3cm}
\section{System and Channel Model}
\begin{figure}{} 
	\centering
	\includegraphics[width=80mm,scale=0.9]{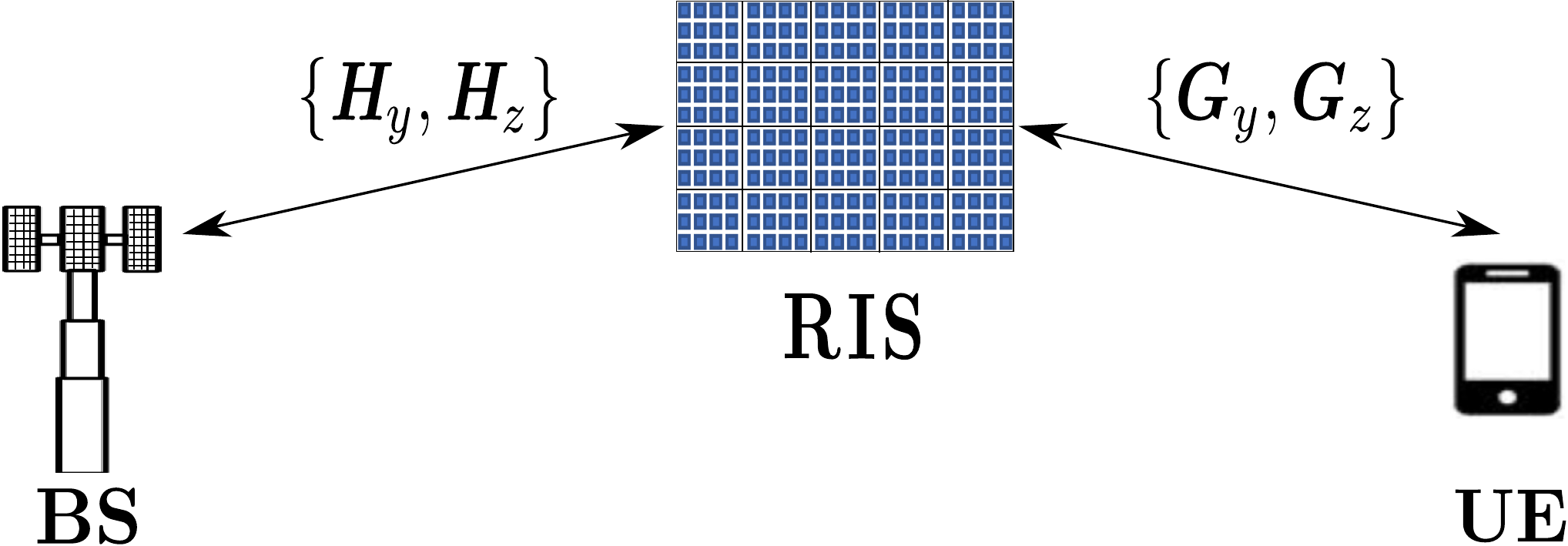}
	\caption{System model.}\label{Fig:1}

\end{figure}

\fea{Let us consider RIS-assisted MIMO communications for a THz scenario, in which we consider the deployment of a URA \cite{Heath_2016}
 placed in the $\text{y}$-$\text{z}$ plane installed at the BS having $M=M_yM_z$ number of transmit antennas, where $M_y$ is the number of antenna elements along the $\text{y}$-axis and $M_z$ is the number of antenna elements along the $\text{z}$-axis. }\textcolor{black}{
 The direct path between the BS and the UE is unavailable due to blockage or deep fade.} Similarly, a URA is deployed at the UE having $Q=Q_yQ_z$ number of receive antenna elements, where $Q_y$ and $Q_z$ denote the antenna elements \gf{along the $\text{y}$ and $\text{z}$ axes}, respectively. Moreover, $N=N_yN_z$ is the number of reflecting elements deployed at the RIS, with $N_y$ and $N_z$ being the number of reflecting elements along the $\text{y}$ and $\text{z}$ axes, respectively (see \figurename~\ref{Fig:1}). The geometrical channel between the BS and the RIS is represented as\fa{,
\begin{equation}
	\bof{H}  =\sum_{r=1}^{R}\alpha_{r}\bof{b}(\phi^r_{\text{ris}_\text{A}},\theta^r_{\text{ris}_\text{A}}) \bof{a}^\R{T}(\phi^r_{\text{bs}},\theta^r_{\text{bs}}) \in \mathbb{C}^{N \times M}, \label{H_sph}
\end{equation}
where $\alpha_{r}$ is the $r$th complex path gain, $\bof{a}(\phi^r_{\text{bs}},\theta^r_{\text{bs}}) \in \mathbb{C}^{M \times 1}$ is the $r$th two-dimensional channel steering vector at the BS, $\phi^r_{\text{bs}}$ is the $r$th azimuth of departure (AoD) and $\theta^r_{\text{bs}}$ as $r$th elevation of departure (EoD).} Similarly, $\bof{b}(\phi^r_{\text{ris}_\text{A}},\theta^r_{\text{ris}_\text{A}}) \in \mathbb{C}^{N \times 1}$ is the $r$th two-dimensional channel steering vector at the RIS, with  $\phi^r_{\text{ris}_\text{A}}$ being the azimuth of arrival (AoA) and $\theta^r_{\text{ris}_\text{A}}$ the elevation of arrival (EoA).
 Analogously, the channel between the RIS and the UE is represented as\fa{ 
\begin{align}
	\bof{G} =\sum_{l=1}^{L}\beta_{l}\bof{q}(\phi^l_{\text{ue}},\theta^l_{\text{ue}}) \bof{p}^\R{T}(\phi^l_{\text{ris}_\text{D}},\theta^l_{\text{ris}_{\text{D}}})  \in \mathbb{C}^{Q \times N}, \label{G_sph}
\end{align}
where $\beta_{l}$ is the $l$th complex path gain, $\bof{p}(\phi^l_{\text{ris}_\text{D}},\theta^l_{\text{ris}_{\text{D}}}) \in \mathbb{C}^{N \times 1}$ is the $l$th two-dimensional channel steering vector at the RIS, with $\phi^l_{\text{ris}_\text{D}}$ and $\theta^l_{\text{ris}_{\text{D}}}$ being the corresponding AoD and EoD, respectively. Similarly, $\bof{q}(\phi^l_{\text{ue}},\theta^l_{\text{ue}}) \in \mathbb{C}^{Q \times 1}$ is the $l$th two-dimensional channel steering vector at the UE, with $\phi^l_{\text{ue}}$ and $\theta^l_{\text{ue}}$ being the associated directional angular parameters.}

The BS array response in $\text{y}$-$\text{z}$ plane can be written as
\begin{equation}
	[\bof{a}(\phi^r_{\text{bs}},\theta^r_{\text{bs}})]_m = e^{-j\pi[(m_y-1)\sin\theta^r_{\text{bs}}\sin\phi^r_{\text{bs}} + (m_z-1)\cos\theta^r_{\text{bs}}]}, \label{ch_3_y-z}
\end{equation}
where $m = m_z + (m_y-1)M_z$, $m_y \in \{1,\dots,M_y\}$, and  $m_z \in \{1,\dots,M_z\}$, as explained in \cite{Asim_2021}. Defining the spatial frequencies as $\mu^r_{\text{bs}} = \pi\sin\theta^r_{\text{bs}}\sin\phi^r_{\text{bs}}$ and $\psi^r_{\text{bs}} = \pi\cos\theta^r_{\text{bs}}$, the channel steering vector can be expressed as the Kronecker product between two channel steering vectors, as 
\begin{equation}
	\bof{a}(\mu^r_{\text{bs}},\psi^r_{\text{bs}}) = \bof{a}_\text{y}(\mu^r_{\text{bs}}) \otimes \bof{a}_\text{z}(\psi^r_{\text{bs}}) \in \mathbb{C}^{M \times 1}, \label{ch_3_kron_channel}
\end{equation}
where 
\begin{align*}
    \bof{a}_\text{y}(\mu^r_{\text{bs}}) &= \left[1, e^{-j\mu^r_{\text{bs}}} ,\dots, e^{-j(M_y-1)\mu^r_{\text{bs}}}\right]^\R{T} \in \mathbb{C}^{M_y \times 1} \\
    \bof{a}_\text{z}(\psi^r_{\text{bs}}) &= \left[1, e^{-j\psi^r_{\text{bs}}} ,\dots, e^{-j(M_z-1)\psi^r_{\text{bs}}}\right]^\R{T} \in \mathbb{C}^{M_z \times 1}.
\end{align*}
In a similar fashion, the channel steering vector $\bof{b}(\phi^r_{\text{ris}_\text{A}},\theta^r_{\text{ris}_\text{A}})$ can also be factorized as the Kronecker product of $\bof{b}_\text{y}(\mu^r_{\text{ris}_\text{A}})$ and $\bof{b}_\text{z}(\psi^r_{\text{ris}_\text{A}})$, respectively.  
{\color{black} The involved channel components $\bof{H}$ and $\bof{G}$ mentioned implicitly in \eqref{H_sph} and  \eqref{G_sph}, respectively, can be expanded in terms of their respective spatial frequencies and Kronecker products as  
	\begin{align}
		\bof{H} \!&=\! \sum_r^R \alpha_{r}\left[\bof{b}_\text{y}(\mu^r_{\text{ris}_\text{A}}) \otimes \bof{b}_\text{z}(\psi^r_{\text{ris}_\text{A}}) \right] \!\! \left[\bof{a}_\text{y}(\mu^r_{\text{bs}}) \!\otimes\! \bof{a}_\text{z}(\psi^r_{\text{bs}})\right]^\R{
			T}, \label{H_spat}\\
		\bof{G} \!&=\! \sum_{l}^{L} \beta_{l}\left[\bof{q}_\text{y}(\mu^l_{\text{ue}}) \!\otimes\! \bof{q}_\text{z}(\psi^l_{\text{ue}}) \right]\!\! \left[\bof{p}_\text{y}(\mu^l_{\text{ris}_\text{D}}) \!\otimes\! \bof{p}_\text{z}(\psi^l_{\text{ris}_\text{D}})\right]^\R{
			T}. \label{G_spat}
	\end{align}

Using the properties $\left(\bof{A} \otimes \bof{B}\right)^\R{T} = \bof{A}^\R{T} \otimes \bof{B}^\R{T}$ and  $\left(\bof{A} \otimes \bof{B}\right) \left(\bof{C} \otimes \bof{D}\right) = \left(\bof{AC} \otimes \bof{BD}\right)$, {\color{black} the rank-one channel matrices $\bof{H}$ and $\bof{G}$ defined in \eqref{H_spat}-\eqref{G_spat} can be reformulated as a Kronecker product of their corresponding $\text{y}$ and $\text{z}$ domain components as 
\begin{align}
	\bof{H} \!&=\! \sum_{r=1}^{R}\underbrace{\alpha_{r}\left[\bof{b}_\text{y}(\mu^r_{\text{ris}_\text{A}})   \bof{a}_\text{y}^\R{
			T}(\mu^r_{\text{bs}})\right]}_{\bof{H}_\text{y}} \otimes \underbrace{\left[\bof{b}_\text{z}(\psi^r_{\text{ris}_\text{A}}) \bof{a}_\text{z}^\R{
			T}(\psi^r_{\text{bs}})\right]}_{\bof{H}_\text{z}}, \label{final_H}\\
   \bof{G} &= \sum_{l=1}^{L}\underbrace{\beta_{l}\left[\bof{q}_\text{y}(\mu^l_{\text{ue}})\bof{p}_\text{y}^\R{
			T}(\mu^l_{\text{ris}_\text{D}})   \right]}_{\bof{G}_\text{y}} \otimes \underbrace{\left[\bof{q}_\text{z}(\psi^l_{\text{ue}}) \bof{p}_\text{z}^\R{
			T}(\psi^l_{\text{ris}_\text{D}})\right]}_{\bof{G}_\text{z}}.
	\label{final_G}
\end{align}
}
\fea{Note that decomposing the channels into their $y$th and $z$th components allows us to exploit their intrinsic rank-one components, which is how we formulate our rank-one approximation problem.} \fa{Under the assumption of THz propagation \cite{Molisch_2021}, the channels consist of a dominant line-of-sight (LOS) component and weaker non-LOS (NLOS) components. When the NLOS terms are negligible, the involved MIMO channels are approximately rank-one matrices, given as $\bof{H} \approx \bof{H}_\text{y} \otimes \bof{H}_\text{z}$, where $\bof{H}_\text{y} \in \mathbb{C}^{N_y \times M_y}$, and $\bof{H}_\text{z} \in \mathbb{C}^{N_z \times M_z}$ are horizontal and vertical domain channels. Similarly, $\bof{G} \approx \bof{G}_\text{y} \otimes \bof{G}_\text{z}$, where $\bof{G}_\text{y} \in \mathbb{C}^{Q_y \times N_y}$ and $\bof{G}_\text{z} \in \mathbb{C}^{Q_z \times N_z}$ are the associated horizontal and vertical component matrices.}
}
The received pilot sequence at the UE \textit{via} RIS by assigning the $k$-th phase-shift pattern at the RIS is represented as
	\begin{equation}
		\bof{X}_{k} = \bof{G} \R{diag}\left\{\bof{\omega}_{k}\right\}\bof{H}\bof{S} + \bof{V}_k \in \mathbb{C}^{Q \times T}, 
		\label{recived_eq}
	\end{equation}
\fa{where $k=1,\dots,K$, is the associated received pilot block, while $\bof{S} \in \mathbb{C}^{M \times T}$ denotes the pilot symbol matrix holding the length-$T$ pilot sequences. The diagonal matrix $\R{diag}\left\{\bof{\omega}_k \right\}
\in \mathbb{C}^{N \times N}$ holds the RIS reflection coefficients ($k$-th column of the discrete Fourier transform (DFT) matrix) used at the $k$-th block  for probing
the propagation channel. It is further assumed that $\bof{H}$ and $\bof{G}$ remain constant during the channel probing stage.}  

	\vspace*{-0.3cm}
\begin{figure}[!t]
	\centering
	\includegraphics[width=80mm,scale=0.9]{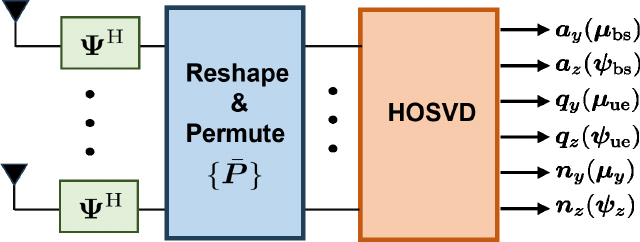}
	\caption{HDR Receiver.}\label{Fig:ab}

\end{figure}
\section{Higher-dimensional Rank-One Approximations (HDR) Method}
{\color{black} Our idea is to exploit the Kronecker decomposition of BS-RIS and RIS-UE channels given in (\ref{final_H}) and (\ref{final_G}) to solve the parametric channel estimation as a whole, using a higher-order rank-one tensor approximation problem. {\color{black} For the convenience of presentation, in the following derivation steps, we focus on the signal part of \eqref{recived_eq} by excluding the noise term (which will be added later). Let $\overline{\bof{X}}_{k} = \bof{G} \R{diag}\left\{\bof{\omega}_{k}\right\}\bof{H}\bof{S}$, so that $\bof{X}_{k}=\overline{\bof{X}}_{k} + \bof{V}_k$. Using the properties $\R{vec}\left\{\bof{ABC}\right\} = (\bof{C}^\R{T} \otimes \bof{A})\R{vec}\left\{\bof{B}\right\}$, $\R{vec}\left\{\bof{A}\R{diag}\left\{\bof{d}\right\}\bof{B}\right\}=(\bof{B}^{\R{T}} \diamond \bof{A})\bof{d}$, and applying the $\R{vec}\{.\}$ operator to $\overline{\bof{X}}_{k}$ leads to
\begin{align}
    \overline{\bof{x}}_{k} \doteq \R{vec}\{\overline{\bof{X}}_{k}\}= \left(\bof{S}^\R{T} \otimes \bof{I}_Q\right) \left( \bof{H}^\R{T} \diamond \bof{G} \right) \bof{\omega}_{k} \in \mathbb{C}^{QT \times 1}.\nonumber 
\end{align}
\textcolor{black}{Defining the matrix $\bof{X} = \left[\overline{\bof{x}}_1,\dots, \overline{\bof{x}}_K\right] \in \mathbb{C}^{QT \times K}$ that collects the resulting signal over $K$ received blocks yields}
	\begin{align}
		\bof{X} =  \big(\bof{S}^\R{T} \otimes \bof{I}_Q\big)\big(\bof{H}^\R{T} \diamond \bof{G}\big)\bof{\Omega}, \label{Xmatrix}
	\end{align}
\fa{where $\bof{\Omega}=[\bof{\omega}_{1},\ldots, \bof{\omega}_{K}] \in \mathbb{C}^{N \times K}$. Recasting (\ref{Xmatrix}) as tensor  
\begin{align}
		\mathcal{X} =   \mathcal{T}_{\mathbf{G},\mathbf{H}} \times_1 \bof{I}_Q \times_2 \bof{S}^\R{T} \times_3 \bof{\Omega}^\R{T} \label{Xtensor},
\end{align}
where $\mathcal{X} \in \mathbb{C}^{Q\times T \times K}$ is the pilot signal tensor, and   $\mathcal{T}_{\mathbf{G},\mathbf{H}} \in \mathbb{C}^{Q \times M \times N}$ is the composite channel tensor expressed as
\begin{align}
	\mathcal{T}_{\mathbf{G},\mathbf{H}} = \mathcal{I}_{3,N }\times_1 \bof{G} \times_2 \bof{H}^\R{T} \times_3 \bof{I}_N.
\end{align}
Using the structure in $\mathcal{T}_{\mathbf{G},\mathbf{H}}$, the 1-mode unfolding of the noiseless pilot signal tensor $\mathcal{X}$ defined in  (\ref{Xtensor}) can be factorized}
\begin{align}
	[\mathcal{X}]_{(1)} = [\mathcal{T}_{\mathbf{G},\mathbf{H}}]_{(1)}(\bof{\Omega} \otimes \bof{S})= \bof{G}(\bof{I}_N \diamond \bof{H}^\R{T})(\bof{\Omega} \otimes \bof{S})\label{Xtensor_1mode}.
\end{align}
\fa{A PARAFAC tensor model is proposed in \cite{Gilderlan_2021} for unstructured channels.} Let $\bof{\Psi} \doteq \bof{\Omega}\otimes \bof{S}\in \mathbb{C}^{MN \times TK}$ be the \textit{effective joint training matrix} that combines the pilot and RIS phase shifts matrices. We design the pilot sequence length $T$ and the RIS training block number $K$ such that $\bof{\Psi}$ is row-wise orthonormal satisfying $\bof{\Psi}\bof{\Psi}^\R{H} = \bof{I}_{MN}$. Several choices ensuring this property are possible, such as a (possibly truncated) $TK\times M$ DFT or Hadamard matrix. Note that such a joint design of the pilot and phase shift matrices requires $TK\geq MN$.}
\footnote{\textcolor{black}{The joint design of $\bof{\Psi}=\bof{\Omega} \otimes \bof{S}$ gives us the flexibility to trade off the pilot sequence length and the number of RIS training blocks to meet the desired orthogonality. For instance, increasing $T$ and decreasing $K$ accordingly yields a finer time delay resolution and a ``wider'' spatial probing. Alternatively, we could also split the joint filtering into two sequential filtering steps, which corresponds to $\mathcal{E}=\tilde{\mathcal{X}} \times_2 \bof{S}^{\ast} \times_3 \bof{\Omega}^{\ast}$. Although this approach is a bit less complex, it would require satisfying both $K\geq N$ and $T\geq M$, which is a more restrictive condition than the joint design.}}
Now, adding the noise tensor $\mathcal{V} \in \mathbb{C}^{Q\times T \times K}$, we have 
\begin{align}
	[\tilde{\mathcal{X}}]_{(1)}= \bof{G}(\bof{I}_N \diamond \bof{H}^\R{T})\bof{\Psi} + [\mathcal{V}]_{(1)}\label{Xtensor_1mode},
\end{align}
Applying right-filtering on \eqref{Xtensor_1mode} using the known joint pilot-RIS training matrix leads to 
\begin{align}
	[\mathcal{E}]_{(1)}\doteq [\tilde{\mathcal{X}}]_{(1)}\bof{\Psi}^\R{H} \approx \bof{G}(\bof{I}_N \diamond \bof{H}^\R{T}) 
 \label{Etensor_1mode},
\end{align}
where $\mathcal{E} \approx \bof{I}_{3,N} \times_1 \bof{G} \times_2 \bof{H}^\R{T} \times_3 \bof{I}_N \in \mathbb{C}^{Q \times M \times N}$. Let 
\begin{align}
\bof{E}\doteq [\mathcal{E}]^\R{T}_{(3)}  \approx \bof{H}^\R{T} \diamond \bof{G} \in \mathbb{C}^{QM \times N}
\label{eq:E}
\end{align}

\begin{algorithm}[!t]
	\caption{\label{alg:HDR} HDR method}
	\Begin{
	Matched filtering: $[\mathcal{E}]_{(1)}\doteq [\tilde{\mathcal{X}}]_{(1)}\bof{\Psi}^\R{H}$ \\
Reshaping: $\bof{E}= \R{reshape}_{QM \times N}\{[\mathcal{E}]_{(1)}\} $\\
Permutation: $\overline{\bof{z}}=\overline{\bof{P}}\bof{E}$ \\ 
Tensorization: $\overline{\mathcal{Z}}=\textrm{tens}(\overline{\bof{z}})  \in \mathbb{C}^{Q_z \times M_z \times N_z \times Q_y \times M_y \times N_y}$ \\ 
$\hat{\bof{u}} \leftarrow \R{HOSVD}\left(\overline{\mathcal{Z}}\right)$ \\ \quad where $\hat{\bof{u}} \in \left\{\hat{\bof{a}}_\text{y},\hat{\bof{a}}_\text{z},\hat{\bof{q}}_\text{y},\hat{\bof{q}}_\text{z},\hat{\bof{n}}_\text{y},\hat{\bof{n}}_\text{z}\right\}$
}
\end{algorithm}

\noindent be an LS approximation to the combined BS-RIS-UE ``Khatri-Rao channel''. It was shown in \cite{Gilderlan_2021} that unstructured estimates of the individual channel matrices $\bof{H}$ and $\bof{G}$ can be obtained from the filtered signal in  (\ref{Etensor_1mode}) by solving the Khatri-Rao factorization (KRF) problem 
\begin{align}
    \hat{\bof{H}},\hat{\bof{G}} = \arg\min\left\| \bof{E} - \bof{H}^\R{T} \diamond \bof{G} \right\|_\R{F}^2.
\end{align}
 Herein, we extend this idea by exploiting the channel's geometric structure employing the Kronecker decomposition of BS-RIS and RIS-UE channels into their respective $\text{y}$ and $\text{z}$ components given in (\ref{final_H}) and (\ref{final_G}), which leads to the following mixed Khatri-Rao-Kronecker factorization problem }
	\begin{align}
  \arg\min\left\| \bof{E} - \left(\bof{H}_\text{y} \otimes \bof{H}_\text{z}\right)^\R{T} \diamond \left(\bof{G}_\text{y} \otimes \bof{G}_\text{z}\right) \right\|_\R{F}^2. \label{E_cost_fuc}
	\end{align}
	{\color{black} Interestingly, we arrive at an equivalent problem by resorting to the property $\left(\bof{A} \otimes \bof{B}\right) \diamond \left(\bof{C} \otimes \bof{D}\right) = \bof{P}_1 \left[\left(\bof{A} \diamond \bof{C}\right) \otimes \left(\bof{B} \diamond \bof{D}\right)\right]$, where $\bof{P}_1 \in \mathbb{R}^{Q_zM_zQ_yM_y  \times Q_zQ_yM_zM_y}$ is a block permutation matrix the structure of which is derived in Appendix \ref{Appendix_permut}. Applying this property and defining $\bof{Z} \doteq \bof{P}_1\bof{E}$ allows us to rewrite (\ref{E_cost_fuc}) as a Kronecker factorization problem in terms of $\text{y}$ and $\text{z}$ Khatri-Rao channels } 
	\begin{align}
  \arg\min\left\| \bof{Z} - (\bof{H}_\text{y}^\R{T} \diamond \bof{G}_\text{y}) \otimes (\bof{H}_\text{z}^\R{T} \diamond \bof{G}_\text{z}) \right\|_\R{F}^2. \label{new_cost_function}
  \vspace{-1ex}
	\end{align}
{\color{black} Let $\bof{z}\doteq \R{vec}\left\{\bof{Z}\right\} \in \mathbb{C}^{Q_zM_zQ_yM_yN_zN_y \times 1}$ and consider the property $\R{vec}\{\bof{A} \} \otimes \R{vec}\{\bof{B}\} = \bof{P}_2\left(\R{vec}\{\bof{A} \otimes \bof{B}\}\right) $, where $\bof{P}_2 \in \mathbb{R}^{Q_zM_zN_zQ_yM_yN_y \times Q_zM_zQ_yM_yN_zN_y}$ is a block permutation matrix, the structure of which is provided in the Appendix \ref{Appendix_permut}. Defining $\overline{\bof{z}}\doteq \bof{P}_2\bof{z} \in \mathbb{C}^{ Q_zM_zN_zQ_yM_yN_y \times 1}$ allows to recast problem (\ref{new_cost_function})  in a vector form as}
\begin{align}
		&\arg\min\left\| \overline{\bof{z}} - \R{vec}\big\{\bof{H}_\text{y}^\R{T} \diamond \bof{G}_\text{y}\big\} \otimes \R{vec}\big\{\bof{H}_\text{z}^\R{T} \diamond \bof{G}_\text{z}\big\} \right\|_\R{F}^2. \label{cost_hdr}
	\end{align}
{\color{black} Note that  $\R{vec}\big\{\bof{H}_\text{y}^\R{T} \diamond \bof{G}_\text{y}\big\} = \bof{n}_\text{y}(\mu_{\text{y}}) \otimes \bof{a}_\text{y}(\mu_{\text{bs}}) \otimes  \bof{q}_\text{y}(\mu_{\text{ue}})$ and $\R{vec}\big\{\bof{H}_\text{z}^\R{T} \diamond \bof{G}_\text{z}\big\} = \bof{n}_\text{z}(\psi_{\text{z}}) \otimes \bof{a}_\text{z}(\psi_{\text{bs}}) \otimes  \bof{q}_\text{z}(\psi_{\text{ue}})$, and define the sixth-order tensor $\overline{\mathcal{Z}} \in \mathbb{C}^{Q_z \times M_z \times N_z \times Q_y \times M_y \times N_y}$ obtained by reshaping the vector $\overline{\bof{z}} \in \mathbb{C}^{Q_zM_zN_zQ_yM_yN_y \times 1}$ accordingly. \fea{Such vector-tensor data mapping is represented by the following ``tensorization'' operator $\overline{\mathcal{Z}}=\textrm{tens}(\overline{\bof{z}})$  that rearranges the entries of the one-dimensional vector as those of a sixth-order tensor. 
Hence, problem \eqref{cost_hdr} is equivalent to a rank-one sixth-order tensor approximation problem  }
}
\begin{align}
&\hat{\bof{a}}_\text{y},\hat{\bof{a}}_\text{z},\hat{\bof{q}}_\text{y},\hat{\bof{q}}_\text{z},\hat{\bof{n}}_\text{y},\hat{\bof{n}}_\text{z} = \arg\min_{\bof{}} \label{hdr_final} \\
		& \left \|\overline{\mathcal{Z}} \!-  \!\bof{q}_\text{z}(\psi_{\text{ue}}) \!\circ\! \bof{a}_\text{z}(\psi_{\text{bs}}) \!\circ\! \bof{n}_\text{z}(\psi_{\text{z}}) 
		\!\circ\! \bof{q}_\text{y}(\mu_{\text{ue}}) \!\circ\! \bof{a}_\text{y}(\mu_{\text{bs}}) \!\circ\! \bof{n}_\text{y}(\mu_{\text{y}})	\right\|_\R{F}^2,   \nonumber  
	\end{align}	
where $\bof{n}_\text{y}(\mu_{\text{y}}) = \bof{b}_\text{y}(\mu_{\text{ris}_\text{A}})\odot\bof{p}_\text{y}(\mu_{\text{ris}_\text{D}})\in\mathbb{C}^{Ny \times 1}$ and $\bof{n}_\text{z}(\psi_{\text{z}}) = \bof{b}_\text{z}(\psi_{\text{ris}_\text{A}})\odot\bof{p}_\text{z}(\psi_{\text{ris}_\text{D}}) \in\mathbb{C}^{Nz \times 1}$ are the effective $\text{y}$-th and $\text{z}$-th domains RIS steering vectors.
  The solution to the problem (\ref{hdr_final}) can be obtained by resorting to the HOSVD algorithm, which in our case consists of six independent rank-one matrix approximations to each matrix unfolding of the tensor $\overline{\mathcal{Z}}$ ({\color{black} due to limited space, we refer the reader to \cite{DeLathauwer2000} for further details on the HOSVD steps}). This problem can also be solved in an iterative way using the higher-order power method \cite{Regalia2000}, which has also been exploited in \cite{Asim_2020} to decode multi-linear Kronecker-structured constellations. It is worth noting that the proposed method enjoys parallel processing since all the six rank-one approximation steps of the HOSVD algorithm can be executed in parallel, thus delivering fast estimates of the involved steering vectors, which is important in scenarios with low latency requirements. Finally, the combined complex path gain $\gamma=\alpha\beta$, i.e., 
 \begin{align}
     \hat{\gamma} = \hat{\mathcal{E}} \times_1 \hat{\bof{q}}^\R{H} \times_2 \hat{\bof{a}}^\R{H}  \times_3 \hat{\bof{n}}^\R{H},
 \end{align}
  \fa{where $\hat{\bof{n}} = \frac{1}{N}(\hat{\bof{n}}_\text{y} \otimes \hat{\bof{n}}_\text{z})$, $\hat{\bof{a}} = \frac{1}{M}(\hat{\bof{a}}_\text{y} \otimes \hat{\bof{a}}_\text{z})$, and $\hat{\bof{q}} = \frac{1}{Q}(\hat{\bof{q}}_\text{y} \otimes \hat{\bof{q}}_\text{z})$.}
By taking into account the filtering step in (\ref{Etensor_1mode}), followed by the tensor reshaping $\bof{E}= \R{reshape}_{QM \times N}\{[\tilde{\mathcal{X}}]_{(1)}\bof{\Psi}^\R{H}\}$ leading to (\ref{eq:E}), as well as the two block permutation steps
$\bof{Z} \doteq \bof{P}_1\bof{E}$ and $\overline{\bof{z}}\doteq \bof{P}_2\R{vec}\{\bof{Z}\}$ 
used in the derivations from (\ref{E_cost_fuc})-(\ref{new_cost_function}) and from (\ref{new_cost_function})-(\ref{cost_hdr}), respectively, we can arrive at a simpler sequence of steps representing the proposed receiver processing linking the received pilot signals to the HOSVD input block, which is translated by the input-output relationship $\overline{\bof{z}} = \bof{P}_2 \R{vec}\left\{\bof{P}_1\bof{E}\right\}=
\bof{P}_2 \left(\bof{I}_N \otimes \bof{P}_1 \right)\R{vec}\{\bof{E}\}$, or equivalently, by filtering, reshaping, and block permutation steps modeled as
$$\overline{\bof{z}}= \overline{\bof{P}}\,\R{vec}\left\{\R{reshape}_{QM \times N}\{[\tilde{\mathcal{X}}]_{(1)}\bof{\Psi}^\R{H}\}\right\},$$
where $\overline{\bof{P}}\doteq \bof{P}_2 \left( \bof{I}_N \otimes \bof{P}_1 \right)$ represents an effective 
 $Q_zM_zN_zQ_yM_yN_y \times Q_zQ_yM_zM_yN_zN_y$
 block permutation matrix that properly shuffles the filtered signal, which is then tensorized to feed the HOSVD block. Figure~\ref{Fig:ab} illustrates the block diagram of the receiver processing steps, and a summary of the proposed HDR method is provided in Algorithm \ref{alg:HDR}. 

 Upon completion of the HOSVD stage, an estimate of the  Khatri-Rao channel $\hat{\bof{E}}= \hat{\bof{H}}^\R{T} \diamond \hat{\bof{G}}$ is obtained from the estimated steering vectors by rebuilding the $\text{y}$ and $\text{z}$ channels. The complexity of HDR is dominated by the filtering and HOSVD steps and is given by 
 \begin{align*}
     \mathcal{O}\Big(Q^2MNTK + QMN\left(Q_z+Q_y+M_z+M_y+N_z+N_y\right)\Big).
 \end{align*}

\vspace*{-0.3cm}

\section{Simulation Results}
\begin{figure}[!t]
		\centering
		\subfigure[NMSE performance for rank-one channels.]
		{\includegraphics[width=1.6in]{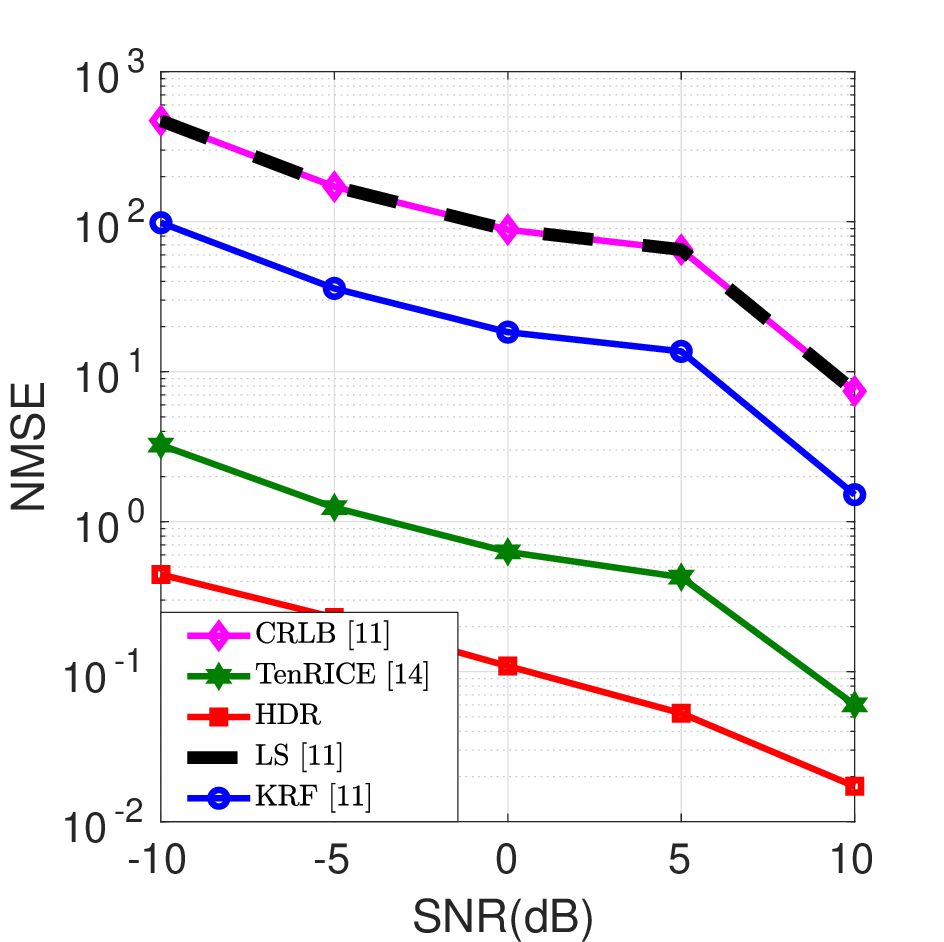}
			\label{fig:9_nmse}
		}
		\subfigure[NMSE performance for higher-rank channels.]
		{
			\includegraphics[width=1.6in]{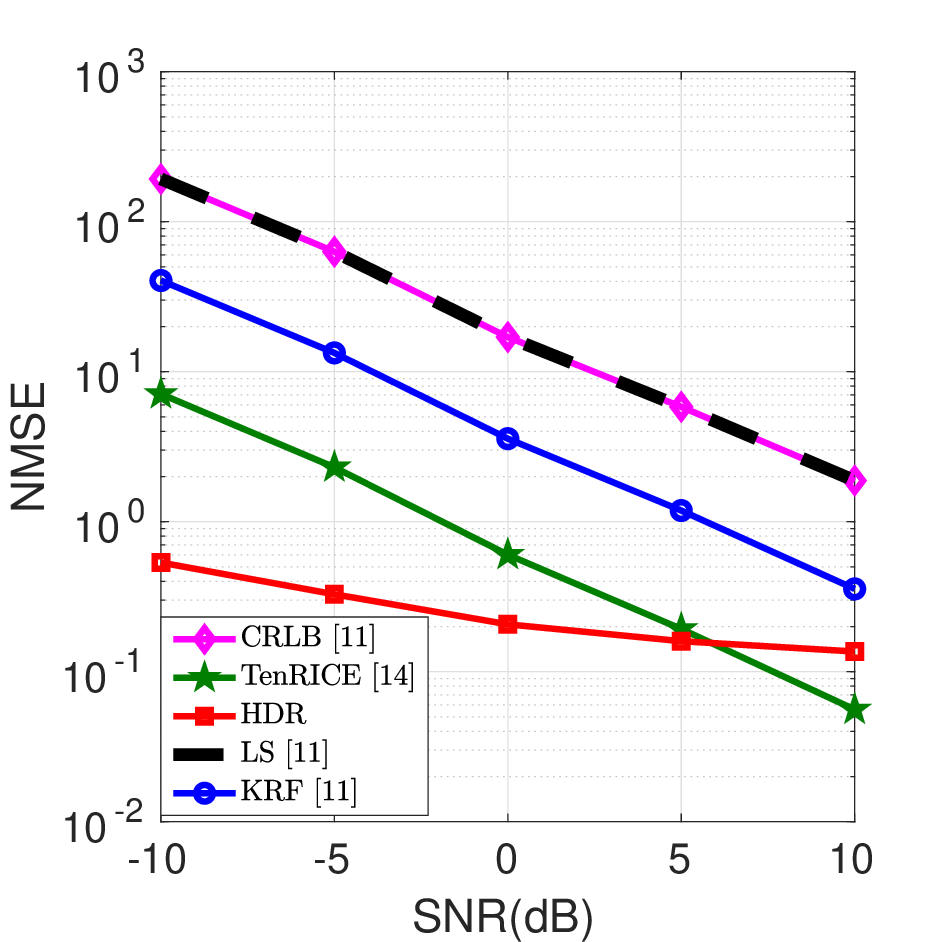}
			\label{fig:10_nmse}
		}
		\caption{NMSE-based performance comparisons.}
		\label{fig:NMSE}
	\end{figure}

\begin{figure}[!t]
		\centering
		\subfigure[SE performance under channel estimation errors assuming multi-rank channels.]
		{\includegraphics[width=1.6in]{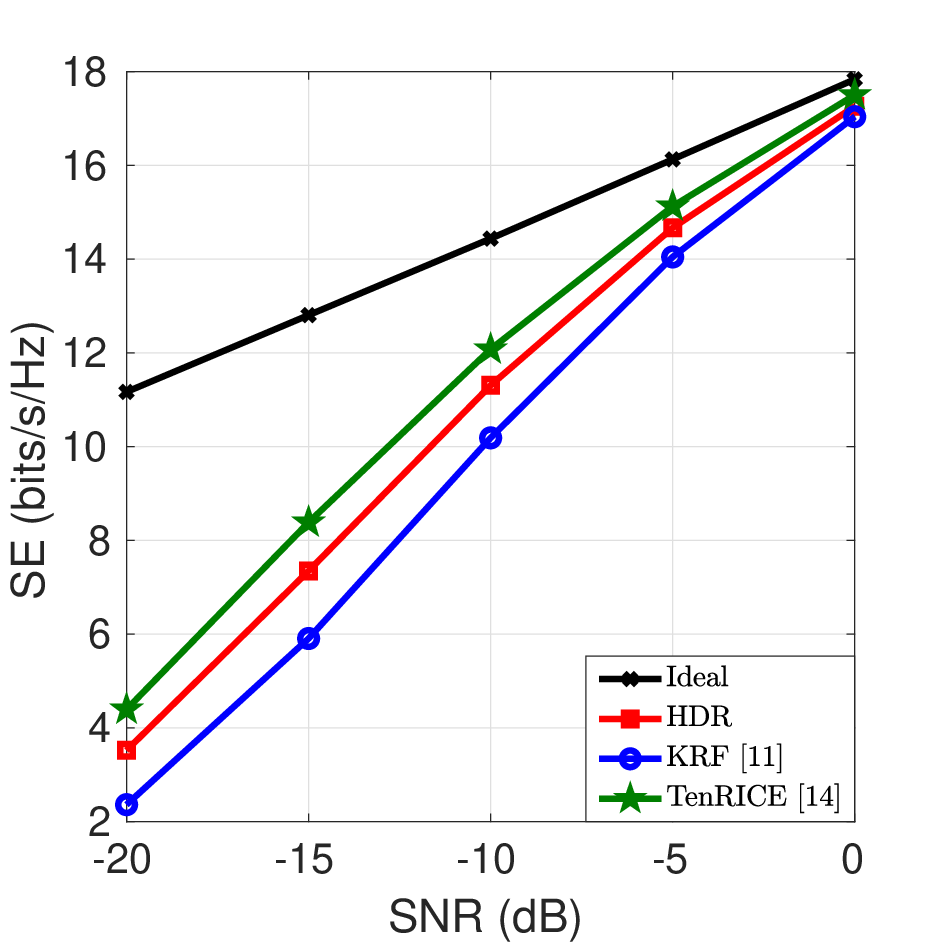}
			\label{fig:11_se}
		}
		\subfigure[Computational Complexity comparison.]
		{
			\includegraphics[width=1.6in]{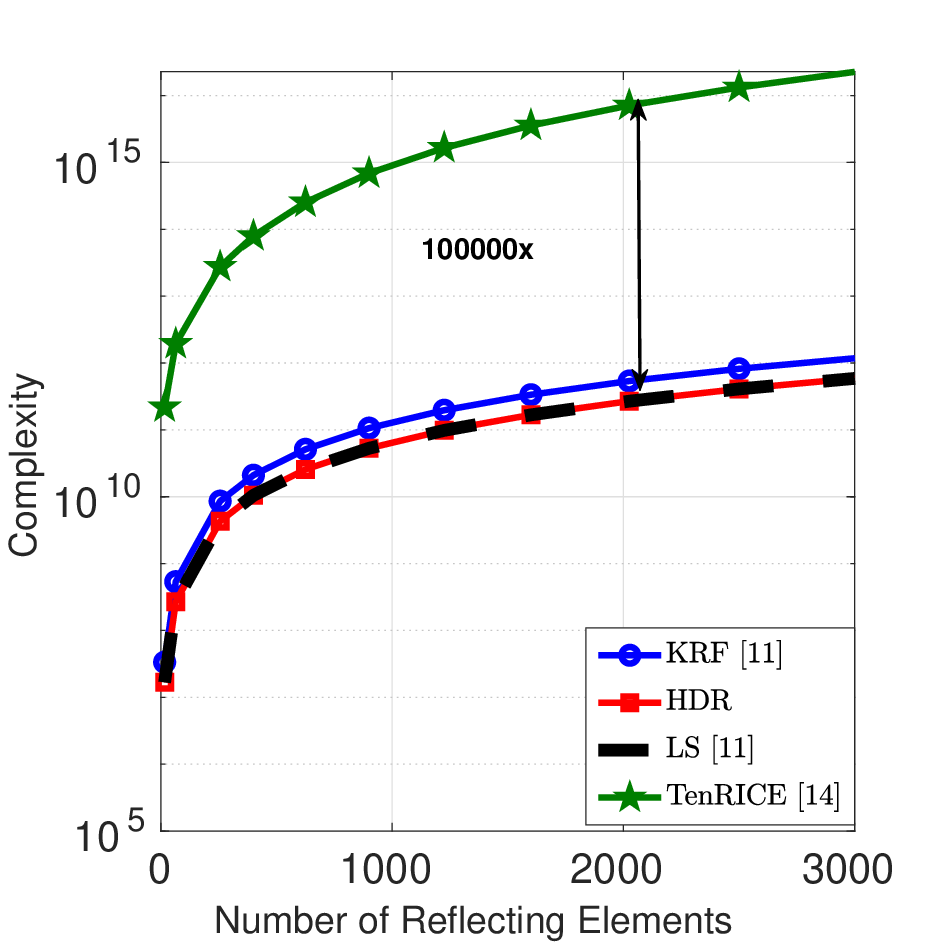}
			\label{fig:a_complexity}
		}
		\caption{SE performance and complexity comparisons.}
		\label{fig:SE_Comp}
	\end{figure}

\fea{We adopt a URA at the BS and the UE, and the channel parameters are modeled and chosen according to \cite{Molisch_2021,Rappaport_2016}}. The AoD $\phi_{\text{bs}}^r$, $\phi_{\text{ris}_\text{D}}^l$ and AoA $\phi_{\text{ris}_\text{A}}^r$, $\phi_{\text{ue}}^l$ are generated with uniform distribution by assuming one sector of a cell as $\phi_{\text{bs}}^r, \phi_{\text{ris}_\text{A}}^r, \phi_{\text{ris}_\text{D}}^l, \phi_{\text{ue}}^l \sim \mathrm{U}\left(-60^\circ,+60^\circ\right)$. EoD $\theta_{\text{bs}}^r$, $\theta_{\text{ris}_\text{D}}^l$ and EoA $\theta_{\text{ris}_\text{A}}^r$, $\theta_{\text{ue}}^l$ are generated by assuming uniform distribution as $\theta_{\text{bs}}^r, \theta_{\text{ris}_\text{A}}^r, \theta_{\text{ris}_\text{D}}^l, \theta_{\text{ue}}^l \sim \mathrm{U}\left(90^\circ,130^\circ\right)$ \cite{Molisch_2021,Rappaport_2016}. \fa{The complex path gains are generated as $\alpha_r,\beta_l \sim \mathcal{CN}(0,1)$. The LOS path is assumed to be stronger than the NLOS paths.} \fea{Note that, for higher-frequency signals, such as mmWave and THz bands, the channel measurement campaigns reveal that the signal power of the LOS component is about 13 dB higher than the sum of the power of NLOS components \cite{Eldeen_muhi2010modelling}.} Moreover, the number of antennas deployed at the BS is $M = 16$, where the number of antennas along the horizontal axis is $M_y=4$ and the number of antennas along the vertical axis is $M_z=4$. Similarly, the number of antennas deployed at the UE is $Q = 16$ having $Q_y=4$ and $Q_z=4$. \fa{Finally, the number of reflecting elements deployed at the RIS is $N=64$}, where $N_y=8$, and $N_z=8$ are fixed for \gf{\figurename~\ref{fig:9_nmse}, to \figurename~\ref{fig:11_se}}. The total transmit power is assumed to be $P_T= \SI{1}{\watt}$ defining $\R{SNR} = P_T/\sigma_n^2$. The  $\R{NMSE}$ for the reconstructed channel is given as 
\begin{align}
    \R{NMSE} =\mathbb{E} \left[ \left\| \bof{E} - \hat{\bof{E}} \right\|^2_F / \left\|\bof{E}\right\|^2_F\right].
\end{align}

\fa{Figure~\ref{fig:9_nmse} evaluates the NMSE-based performance for rank-one channels} of our proposed HDR method in comparison with \fa{the TenRICE \cite{GherekhlooK_2021}, KRF and LS methods \cite{Gilderlan_2021}.} The Khatri-Rao channel given in \eqref{eq:E} is reconstructed using the respective estimations of the channel steering vectors and \fa{combined complex path gain}. \fa{The HDR method outperforms the TenRICE \cite{GherekhlooK_2021}, the KRF, and the LS methods \cite{Gilderlan_2021}.} This is because of exploiting the inherent factorization structure and modeling this as a higher-order tensor to obtain a significant tensor gain. \fa{The TenRICE method \cite{GherekhlooK_2021} partially exploits the geometrical structure while both the LS and KRF methods \cite{Gilderlan_2021} are unable to exploit the geometrical structure of the problem.} The LS method for unstructured channel estimation satisfies the normalized CRLB \cite{Gilderlan_2021}, which is derived for unstructured channels. The HDR method and the TenRICE \cite{GherekhlooK_2021} method outperform the normalized
CRLB due to noise rejection by fully and partially exploiting the geometrical structure of the channels, respectively. 
\gf{On} the other hand, the KRF method \cite{Gilderlan_2021} shows improvement compared with the LS method and the normalized CRLB due to noise rejection involved in the LS Khatri-Rao factorization step and due to the harmless influence of noise on the reconstruction of the factor matrices, respectively.
\fa{The HDR method shows almost 100 times better performance \gf{compared to} the KRF method \cite{Gilderlan_2021}, and 10 times compared to the TenRICE method \cite{GherekhlooK_2021} at \SI{0}{\decibel} SNR.}
\fa{To further analyze the performance of our proposed HDR method in higher-rank channels, Fig.~\ref{fig:10_nmse} shows that the HDR method still presents the best performance in the low SNR regime (-5 dB to 0 dB) but does not show further improvement \fa{after 5 dB SNR} due to the approximation error caused by the assumption of a THz propagation scenario.} 

\fea{The	performance of the proposed HDR algorithm is further evaluated in \figurename~\ref{fig:11_se}, in terms of SE as defined in \cite{Asim_2023} assuming multi-rank channels.} \fa{The active and passive beamforming vectors are calculated using the solution of \cite{Zappone_2021}, where an additional noise rejection is achieved.} Here, the design of the \fa{active and passive beamforming vectors is based on the estimated channels. We also show the SE results for perfectly known channels as a benchmark. The HDR-based channel estimation shows almost similar performance compared to the TenRICE \cite{GherekhlooK_2021} and the KRF methods \cite{Gilderlan_2021} for higher-rank channels. 
The reason behind the minor improvement in the case of the TenRICE method is that the SVD-based precoder and the combiner design are based on estimated multi-rank channels while the HDR method only estimates the dominant path of the channel, and, therefore, the SVD-based precoder and the combiner are based on the approximated rank-one channel, which shows a little degradation in the SE performance.  Still, HDR shows a slightly better SE performance than KRF at SNR=\SI{-15}{\decibel} but a similar performance compared to TenRICE due to the approximation error of higher-rank channels as rank-one channels.}

\fa{Figure \ref{fig:a_complexity} depicts a complexity comparison of our proposed HDR algorithm with the TenRICE \cite{GherekhlooK_2021}, KRF, and LS methods \cite{Gilderlan_2021}.} Exploiting the inherent factorization structure and modeling it as a higher-order tensor help to reduce the overall complexity. \fa{The HDR method has almost the same computational complexity as the LS method \cite{Gilderlan_2021} for different numbers of RIS elements but outperforms the TenRICE \cite{GherekhlooK_2021}, and the KRF \cite{Gilderlan_2021} methods.} 
\vspace*{-0.3cm}

\section{Conclusions}
{\color{black}

We proposed a tensor-based channel estimation method for THz channels, which exploits the inherent geometrical structures of the involved channel matrices to recast the channel parameter estimation as a single sixth-order tensor rank-one approximation problem that can be efficiently solved using the HOSVD. The proposed HDR method outperforms the competing TenRICE, KRF, and LS methods in terms of estimation accuracy with a similar spectral efficiency. Additionally, HDR has a lower complexity than its competitors. 
}

%
\vspace*{-0.3cm}

\appendices

\section{Block permutation matrices $\bof{P}_1$ and $\bof{P}_2$} \label{Appendix_permut}
The two block-permutation matrices used in step~4 of Algorithm 1 can be defined in a unified and general form using Kronecker products of canonical unit vectors as follows 
\begin{equation}
	\bof{P}_r \doteq \sum_{i=1}^{I} \sum_{j=1}^{J} \sum_{k=1}^{K} \sum_{l=1}^{L} \bof{p}^{(i,k,j,l)}_{\text{b}_{(r)}} \bof{p}^{(i,j,k,l)\R{T}}_{\text{a}_{(r)}} \in \mathbb{R}^{IKJL \times IJKL}, 
 \vspace{-2ex}
\end{equation}
$r=1, 2$, where
\vspace{-1ex}
\begin{align}
	\bof{p}^{}_{\text{a}_{(r)}}& =  \bof{e}_l^{(r)} \otimes \bof{e}_k^{(r)} \otimes \bof{e}_j^{(r)} \otimes \bof{e}_i^{(r)}\\
 \bof{p}^{}_{\text{b}_{(r)}}& = \bof{e}_l^{(r)} \otimes \bof{e}_j^{(r)} \otimes \bof{e}_k^{(r)} \otimes \bof{e}_i^{(r)}. 
\end{align}
For $r=1$, we have the following correspondences $\left(I,J,K,L\right)\leftrightarrow \left(Q_z,Q_y,M_z,M_y\right)$, while for $r=2$, we have $\left(I,J,K,L\right)\leftrightarrow \left(Q_zM_z,Q_yM_y,N_z,N_y\right)$.
\ifCLASSOPTIONcaptionsoff
  \newpage
\fi



%

\vspace*{-0.3cm}
\bibliographystyle{IEEEtran}
\bibliography{IEEEabrv,letterbib_wcl}
%
%

%

%
%
%




\end{document}